\documentclass[
 reprint,
 amsmath,amssymb,
 aps,
 pra,
 floatfix,
 superscriptaddress
]{revtex4-2}
\usepackage{graphicx}
\usepackage{caption}
\usepackage{subcaption}
\usepackage{bm}
\usepackage{dcolumn}
\usepackage{pgfplots}
\pgfplotsset{compat=1.18}
\usepackage{tikz}
\usetikzlibrary{arrows.meta, decorations.pathmorphing}
\usepackage{xcolor}

\usepackage{braket}
\usepackage{ulem}

\usepackage[colorlinks]{hyperref}

\begin{document}

\title{Quantum Trajectory Entanglement in Seeded Boundary Time Crystals}

\author{Mohammad Jafari}
\email{mojafari@id.uff.br}
\affiliation{Instituto de F\'isica, Universidade Federal Fluminense, Av. Gal. Milton Tavares de Souza s/n, Gragoat\'a, 24210-346 Niter\'oi, Rio de Janeiro, Brazil}

\author{Fernando Iemini}
\affiliation{Instituto de F\'isica, Universidade Federal Fluminense, Av. Gal. Milton Tavares de Souza s/n, Gragoat\'a, 24210-346 Niter\'oi, Rio de Janeiro, Brazil}

\begin{abstract}
We investigate the entanglement dynamics along quantum trajectories during the seeding of time-crystalline order in a boundary time crystal (BTC). Specifically, how entanglement spreads among different spin ensembles when a BTC attempts to seed its time-crystalline behavior onto otherwise static spin ensembles, through a collective dissipative channel.
We analyse both the dynamical growth of entanglement in time and the steady-state properties of the system. Our results reveal two fundamentally distinct regimes. In the seeded BTC phase, the steady-state entanglement entropy between the ensembles grows with system size $N$, accompanied by macroscopic fluctuations along the trajectories. In contrast, in the non-seeded static phase, both the steady-state entanglement and its fluctuations decay exponentially with $N$. The model thus features a measurement-induced phase transition (MIPT) driven by the seeding mechanism.  Furthermore, these findings establish dissipative seeding as a powerful mechanism for controlling quantum correlations in open many-body systems, with direct experimental relevance  to this class of model without a postselection barrier.

    \end{abstract}

\maketitle

\section{Introduction}

Understanding how correlations emerge and propagate in open quantum many-body systems is a central question in non-equilibrium quantum physics. While dissipation has traditionally been viewed as a source of decoherence that destroys quantum correlations, it is now well established that structured dissipation can instead act as a powerful resource for generating entanglement, synchronization, and non-trivial steady states \cite{fazioManyBody2024,Diehl2008, Verstraete2009, Daley2014}. In particular, collective coupling to a shared environment can induce long-range correlations and give rise to dissipative phase transitions and dynamical ordering phenomena \cite{Sieberer2016, Iemini2018, Hajdusek2022}.

A complementary perspective on open-system dynamics is provided by quantum trajectory methods, where the evolution of the density matrix is unraveled into stochastic realizations of pure states undergoing non-Hermitian evolution interrupted by quantum jumps \cite{Daley2014, Dalibard1992, Plenio1998}. This trajectory-resolved approach allows one to track the build-up of correlations at the level of individual realizations, avoiding the classical mixing introduced by ensemble averaging. As a result, information-theoretic quantities such as entanglement entropy can be used to directly probe genuine quantum correlations generated by the dissipative dynamics, providing access to phenomena that are invisible at the level of the average density matrix \cite{LeGal2024, Passarelli2024}.

A remarkable manifestation of such dissipative many-body phenomena are boundary time crystals (BTCs) \cite{Iemini2018, Krishna2023, Kessler2021, Chen2023, Wu2024, Yang2025, Huang2025, Wang2025, Arumugam2026}. These are open quantum systems in which time-translation symmetry is spontaneously broken due to the competition between coherent driving and collective decay, leading to
a dynamical state characterized by persistent oscillations of observables such as the magnetization.
 This phase has attracted considerable attention not only for its fundamental interest, exhibiting many notable phenomena at the ensemble level such as extensive many-body correlations \cite{Lourenco2022}, gapless Lindbladian excitations \cite{Souza2023}, effective non-Markovian correlation dynamics \cite{Carollo2022}, but also for its potential in emerging quantum applications, mostly in quantum metrology sensors \cite{SciPostPhys.18.3.100,PhysRevLett.132.050801,lee2025timescalessqueezingheisenbergscalings,Li_2025,5gh9nmv8,midha2025metrologyopenquantumsystems,oconnor2026attainingquantumsensingenhancement,Pavlov_2023,arumugamElectricfield2025,montenegroQuantum2023,cabot2026parameterestimationonetwotime,PhysRevA.109.L050203,7m63-lnb8,9h67-kqyz,PhysRevB.111.024315,manya2026optimalobservablesquantumenhancedsensing,MONTENEGRO20251} but also first steps along quantum clocks \cite{singh2025quantumthermodynamicslimitcycle,dj21gmdj}, thermodynamics \cite{Carollo_2024,Paulino_2026} and photonics timetronics \cite{zheludevTime2024}.
An interesting extension of the BTC phenomenology is the concept of dissipative seeding, recently proposed in Ref.~\cite{Hajdusek2022}. In this scenario, a subsystem already exhibiting BTC behavior is coupled to a second, otherwise static subsystem through a shared dissipative channel. The time-crystalline order of the first subsystem can then ``seed'' persistent oscillations into the second, leading to synchronized dynamics across the entire system. This seeding mechanism provides a powerful route for controlling and transferring dynamical order in open many-body systems, with potential applications ranging from quantum synchronization to the engineering of correlated steady states.

Notably, such collective dissipative spin models have recently been shown to exhibit a measurement-induced entanglement transition at the level of individual quantum trajectories without a postselection barrier \cite{Passarelli2024,delmonteMeasurementinduced2025,liEmergent2025}. This is a crucial feature, as conventional monitored quantum systems are severely constrained by the postselection barrier: the probability of reproducing a given quantum trajectory decays exponentially with the number of recorded jump events, making experimental access to trajectory-level observables prohibitively difficult for large systems. In these collective models, however, the entanglement saturation time grows only logarithmically with system size, reducing the postselection overhead from exponential to polynomial scaling and making them highly appealing from an experimental perspective. Given this context, deepening our understanding of the dissipative seeding mechanism - particularly with regard to how it generates, transfers, and stabilizes quantum correlations at the trajectory level - is both timely and compelling. While the seeding of time-crystalline order has been analyzed through collective magnetization observables and synchronization diagnostics at the ensemble level \cite{Hajdusek2022}, its effects on genuine quantum correlations, such as entanglement between the constituent subsystems, remain unexplored.

\begin{figure}
\centering
\includegraphics[width=0.85\columnwidth]{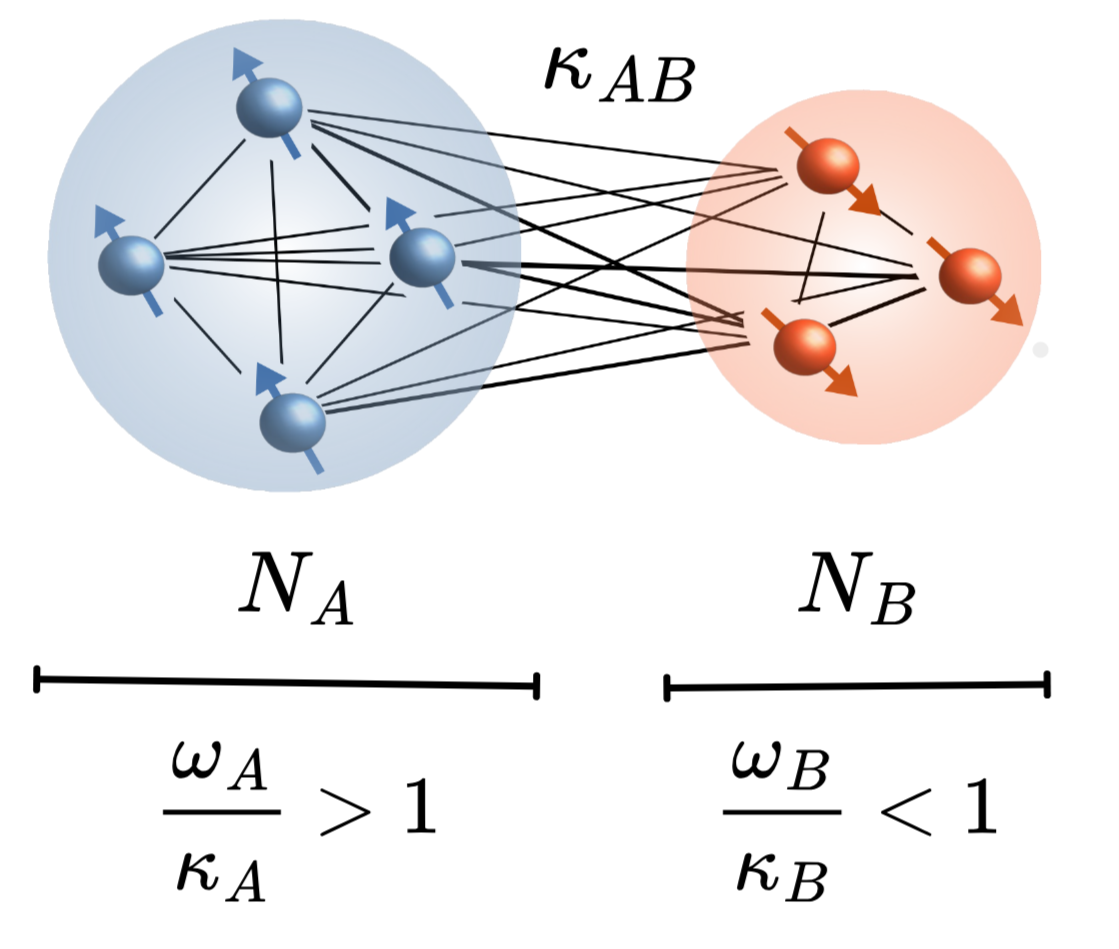}
\includegraphics[width=0.9\columnwidth]{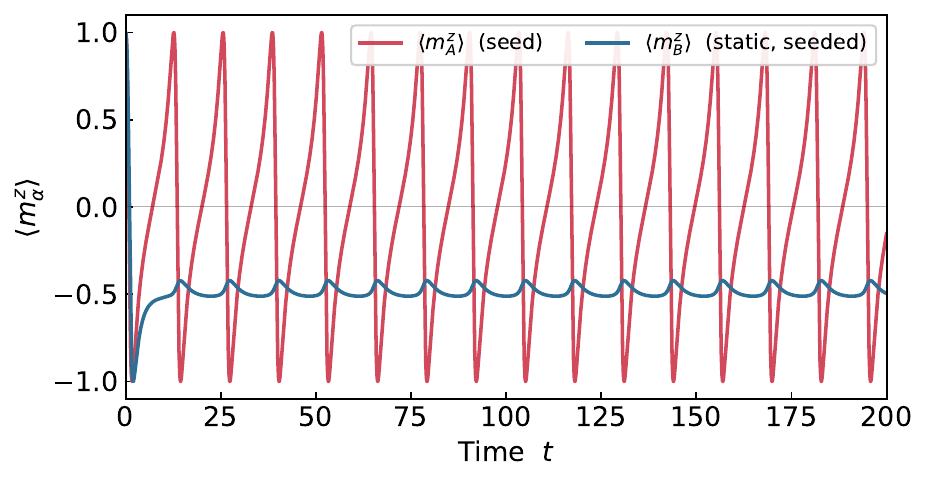}
\caption{ \textbf{(top panel)} Schematic representation of the two-ensemble system. Each ensemble of spins is driven coherently by a magnetic field ($\omega_\alpha$) and is subject to an intra-ensemble collective dissipation ($\kappa_\alpha$), while an inter-ensemble collective decay channel ($\kappa_{AB}$) couples the two ensembles through the environment. We consider the situation where ensemble A is initially in a boundary time-crystal phase ($\omega_A/\kappa_A > 1$), while ensemble B is initially in a static phase ($\omega_B/\kappa_B < 1$). The collective coupling can seed the time crystallisation across the entire system. \textbf{(bottom panel)} We illustrate the seeded magnetization dynamics within mean-field approach,with $\omega_A = 1.2$, $\omega_B = 0.95$, $\kappa_A = \kappa_B = 1.0$,
and $\kappa_{AB} = 0.1$, both ensembles being initialized in the fully
polarized state $m^z_\alpha(0) = 1$.
}
\label{fig:schematic_model}
\end{figure}

In this work we investigate, from a quantum-trajectory perspective, how dissipative seeding generates correlations in a system composed of two spin ensembles. While one ensemble is in a time-crystal phase, the other is initially in a static phase. The two ensembles are then coupled through a collective decay channel that mediates the transfer of information and correlations between them - as shown schematically in Fig.\eqref{fig:schematic_model}. We analyze the inter-ensemble correlations between the two spin ensembles, unraveling both its dynamical growth of correlations in time as well as its steady-state properties.
Our analysis reveals two sharply distinct regimes. When seeding succeeds, the steady-state entanglement between $A$ and $B$, along with its fluctuations, increases with system size $N$. In contrast, when the seeding fails, both quantities die off exponentially as $N$ grows - a clear signature of a measurement-induced phase transition (MIPT). Approaching the transition point, the derivative of the steady-state entanglement with respect to the dissipative coupling strength grows with $N$, indicating a divergence susceptibility in the thermodynamic limit. These results highlight how the seeding of time-crystalline order can fundamentally reshape the scaling of quantum correlations in open many-body systems, and establish dissipative seeding as a powerful mechanism for controlling entanglement in experimentally accessible settings.

This manuscript is organized as follows. In Sec.~\ref{sec:model} we introduce the model and discuss its properties. In Sec.~\ref{sec:trajectories} we present the quantum-trajectory approach used to unravel the dissipative dynamics. In Sec.~\ref{sec:diagnostics} we define the correlation quantifier employed throughout the work. In Sec.~\ref{sec:inter.ensemble} we present our results for the inter-ensemble correlations. Finally, in Sec.~\eqref{sec.conclusion} we summarize our findings, present our conclusions and perspectives.

\section{Model}
\label{sec:model}

In this section we define our model and discuss its main properties. We consider a system composed of two spin-$1/2$ ensembles, denoted $A$ and $B$, each containing $N_\alpha$ spins ($\alpha \in \{A,B\}$). We focus our analysis on the case where $N_A = N_B \equiv N$, although the formulation remains valid for unequal ensemble sizes.

The open-system dynamics of the combined system is described by a Gorini--Kossakowski--Sudarshan--Lindblad master equation as follows \cite{Hajdusek2022},
\begin{equation}
\dot{\hat{\rho}}
=
- i [\hat H,\hat{\rho}]
+
\sum_{\mu}
\hat L_\mu \hat{\rho}\, \hat L_\mu^\dagger
-
\frac{1}{2}
\{\hat L_\mu^\dagger \hat L_\mu,\hat{\rho}\},
\label{eq:lindblad}
\end{equation}
where $\hat H$ is the coherent Hamiltonian and $\{\hat L_\mu\}$ denote the dissipative jump operators.

The coherent part of the dynamics corresponds to a magnetic field along the x-direction acting on each ensemble,
\begin{equation}
\hat H
=
\sum_{\alpha=A,B}
\omega_\alpha \hat S^x_\alpha ,
\label{eq:H}
\end{equation}
where $\omega_{\alpha}$ is the strength of the magnetic field, and $\hat S_\alpha^\beta =
\sum_{j=1}^{N_\alpha}
\hat \sigma_{\alpha,j}^\beta/2 $, with $\beta=x,y,z$, are collective spin operators acting on the $\alpha$'th ensemble, with $\hat \sigma_{\alpha,j}^\beta$ the corresponding Pauli operators for the j'th spin in the ensemble.
 This coherent drive tends to generate collective precession of the spins, competing with the non-commuting dissipative processes.

The system is subject to both a collective intra-ensemble decay and an inter-ensemble one. A schematic representation of the full two-ensemble setup is shown in Fig.~\ref{fig:schematic_model}-(top panel). Specifically, while the intra-ensemble decay acts independently on each ensemble, with
\begin{equation}
\hat L_\alpha = \sqrt{\frac{\kappa_\alpha}{N_\alpha/2}}
\hat S^-_\alpha,
\qquad
\alpha \in \{A,B\},
\label{eq:Llocal}
\end{equation}
the inter-ensemble channel dissipatively couples the two ensembles through the environment, as follows,
\begin{equation}
\hat L_{AB} =
\sqrt{\frac{\kappa_{AB}}{(N_A+N_B)/2}}
\left(\hat S^-_A + \hat S^-_B\right),
\label{eq:L12}
\end{equation}
where the collective lowering operators are defined as
$\hat S^-_\alpha = \hat S^x_\alpha - i \hat S^y_\alpha$.
The parameters $\kappa_{\alpha}$ and $\kappa_{AB}$ control the strength of these collective dissipative channels. The normalization factors proportional to $N_\alpha$ ensure a well-defined thermodynamic scaling of the dissipative rates.

The shared dissipative channel ($\hat L_{AB}$) is the key ingredient behind the dissipative seeding and synchronization phenomenology \cite{Hajdusek2022}. In its absence, the two ensembles exhibit independent dynamical behaviors. However, one may consider that  ensemble $A$ is chosen in the boundary-time-crystal regime, where coherent driving dominates over collective decay ($\omega_A/\kappa_A > 1$) and the collective magnetization exhibits persistent oscillations. Conversely, ensemble $B$ may be set in a static regime ($\omega_B/\kappa_B < 1$) in which, if isolated, its magnetization would relax to a stationary value \cite{Iemini2018}. When the shared decay channel is switched on, $\kappa_{AB}>0$, the oscillatory dynamics of ensemble $A$ can seed oscillations in ensemble $B$, leading to synchronized or frequency-locked magnetization dynamics \cite{Hajdusek2022}.

This phenomenology can be captured from a mean-field analysis of the spin average magnetization dynamics. The approach proceeds by closing the expectation values for the magnetization at the second-order cumulant level, i.e.,
$
\langle \hat S_\alpha^\beta \hat S_{\alpha'}^{\beta'}\rangle = \langle \hat S_\alpha^\beta \rangle \langle \hat S_{\alpha'}^{\beta'} \rangle,
$
and studying the macroscopic magnetizations \( m_\alpha^\beta = \langle \hat S_\alpha^\beta \rangle\) within the Heisenberg picture. The corresponding equations of motion are given by:
\begin{equation}
\begin{aligned}
\dot{m}^{x}_{\alpha} ={}& \Big(\kappa_\alpha+\tfrac{\kappa_{AB}}{2}\Big)
        m^{x}_{\alpha}m^{z}_{\alpha}
        + \tfrac{\kappa_{AB}}{2}\, m^{z}_{\alpha}m^{x}_{\beta},\\[2pt]
\dot{m}^{y}_{\alpha} ={}& -\,\omega_\alpha\, m^{z}_{\alpha}
        + \Big(\kappa_\alpha+\tfrac{\kappa_{AB}}{2}\Big)
        m^{y}_{\alpha}m^{z}_{\alpha}
        + \tfrac{\kappa_{AB}}{2}\, m^{z}_{\alpha}m^{y}_{\beta},\\[2pt]
\dot{m}^{z}_{\alpha} ={}& \omega_\alpha\, m^{y}_{\alpha}
        - \Big(\kappa_\alpha+\tfrac{\kappa_{AB}}{2}\Big)
          \big[(m^{x}_{\alpha})^{2}+(m^{y}_{\alpha})^{2}\big]\\
        &- \tfrac{\kappa_{AB}}{2}\big(m^{x}_{\alpha}m^{x}_{\beta}
          + m^{y}_{\alpha}m^{y}_{\beta}\big),
\end{aligned}
\label{eq:meanfield}
\end{equation}
We illustrate a seeded BTC dynamics in Fig.~\ref{fig:schematic_model}-(bottom panel). It is interesting to remark that, upon increasing the coupling strength \(\kappa_{AB}\), one also observes an increase in the amplitude of the seeded oscillations, indicating a transfer of collective coherence from the oscillatory ensemble to the otherwise static one.

For sufficiently large \(\kappa_{AB}\), however, this same dissipative coupling drives the system towards a strongly damped regime with no time-crystal phase, in which coherent oscillations are suppressed. This critical coupling can be obtained from the mean-field fixed point solutions via a linear stability analysis, and is given by \cite{Hajdusek2022}:
\begin{equation}
\kappa_{\mathrm{AB,crit}}
=
\frac{2\kappa(\omega_A-\kappa)}
{2\kappa-(\omega_A-\omega_B)},
\label{eq:Gamma_crit}
\end{equation}
Above this threshold, the fixed point becomes stable, and the persistent oscillatory phase is destroyed, marking the boundary of the dissipative time-crystalline region in the parameter space.

\subsection{Symmetric subspace representation}

Notice that, due to the collective dissipative nature of the dynamics, each of the ensembles conserves its total angular momentum. Therefore, given that each ensemble is initialized in a permutationally symmetric configuration, the dynamics remain within this symmetry subspace \cite{Shammah2018,Iemini2024}, also known as the Dicke manifold.  The Hilbert space of each ensemble $\alpha$ is therefore spanned by the Dicke states,
 $\{\ket{D_r^{(\alpha)}}\}_{r=0}^{N_\alpha}$,
where $r$ denotes the number of spins in the excited state $\ket{\uparrow}$ within the symmetric Dicke basis. As a result, the effective Hilbert-space dimension of each ensemble is reduced from $2^{N_\alpha}$ to $N_\alpha + 1$, and the global Hilbert space of the system has therefore dimension $(N_A+1)(N_B+1)$. This property allows the simulation of large collective spin systems while preserving the relevant collective degrees of freedom.

\section{Quantum trajectories}
\label{sec:trajectories}

In order to analyze the dynamics generated by the Lindblad master equation (Eq.\eqref{eq:lindblad}), we employ a quantum trajectory (Monte Carlo wave-function) approach \cite{Dalibard1992,Plenio1998,Daley2014}. In this framework the open-system dynamics is unraveled into stochastic realizations of pure states evolving under non-Hermitian dynamics interrupted by random quantum jumps.
 As a result, information-theoretic quantities such as entanglement entropy and mutual information can be evaluated at the level of individual realizations, allowing us to characterize the build-up of genuine quantum correlations generated by the dissipative dynamics.

\begin{figure*}
\begin{subfigure}{0.325\linewidth}
\centering
\includegraphics[width=\linewidth]{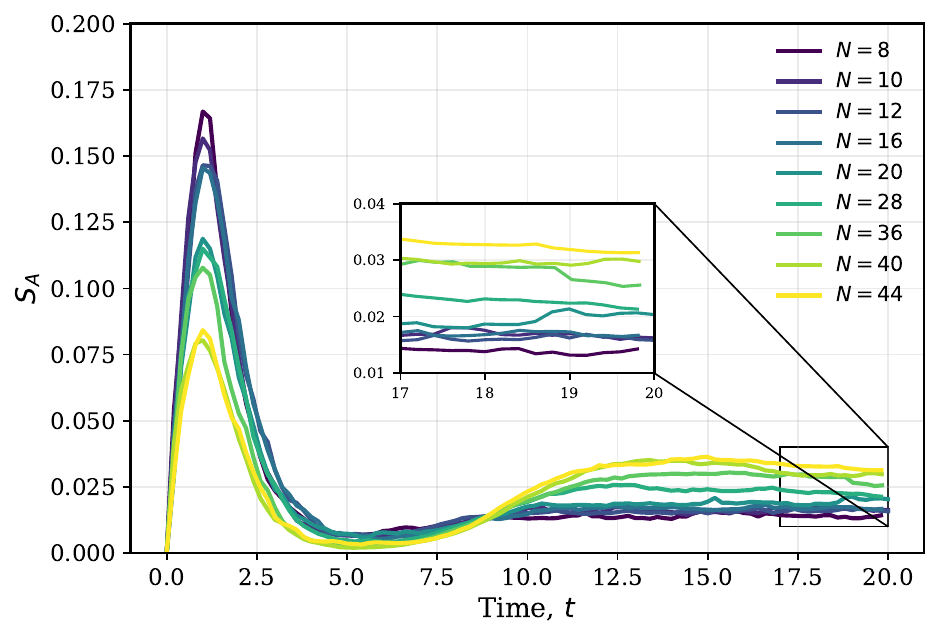}
\caption{$\kappa_{AB}=0.1$}
\end{subfigure}
\hfill
\begin{subfigure}{0.325\linewidth}
\centering
\includegraphics[width=\linewidth]{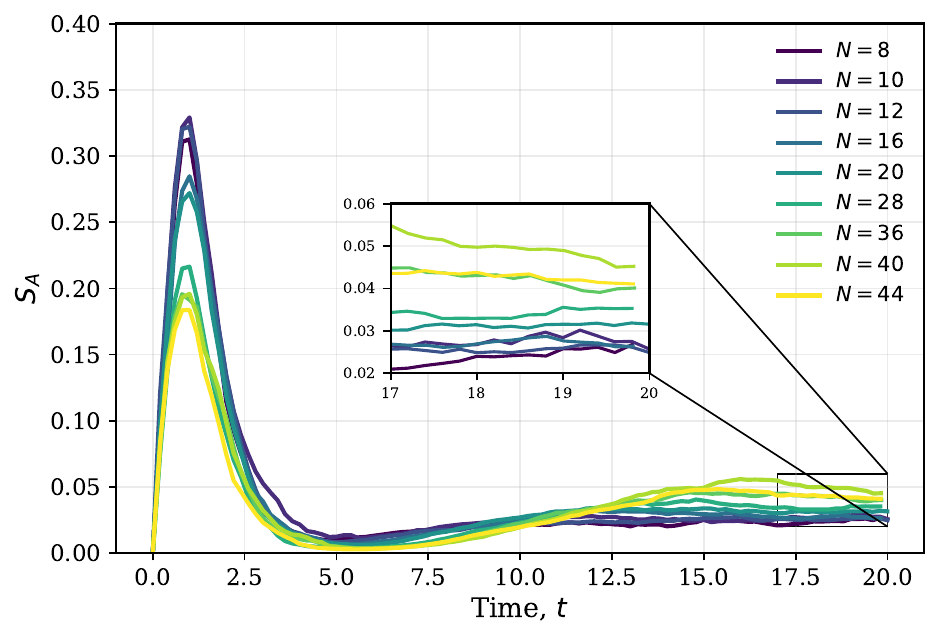}
\caption{$\kappa_{AB}=0.22$}
\end{subfigure}
\hfill
\begin{subfigure}{0.325\linewidth}
\centering
\includegraphics[width=\linewidth]{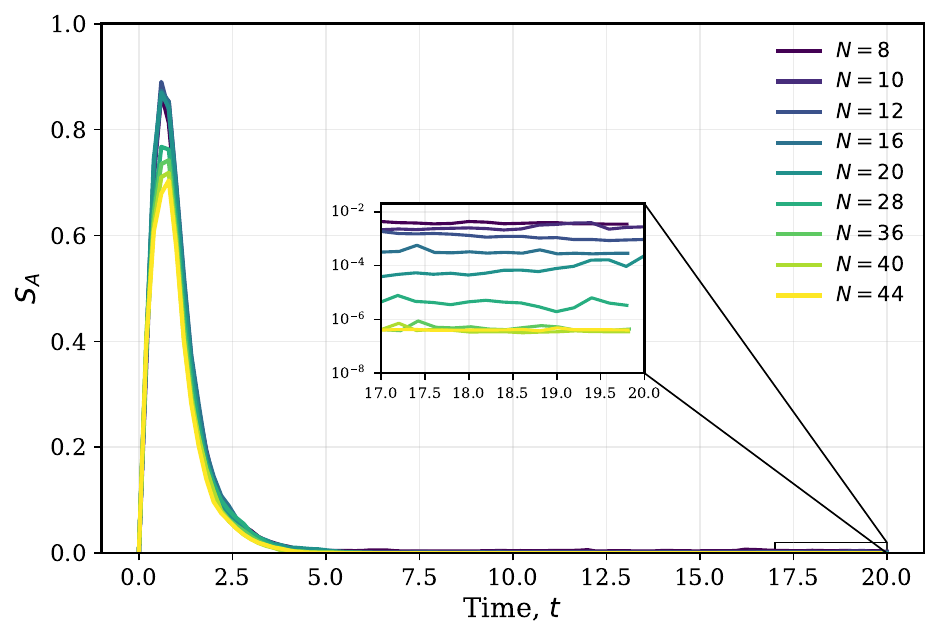}
\caption{$\kappa_{AB}=1.0$}
\end{subfigure}
\caption{
Dynamics of inter-ensemble entanglement entropy $S_A(t)$ for different values of the collective decay strength $\kappa_{AB}$ and varying system sizes $N$. The dynamics is characterized by an initial growth of correlations, followed by a peak and a relaxation toward a steady-state plateau, whose value depends on the strength of the collective decay.
}
\label{fig:entanglement_dynamics}
\end{figure*}

The approach proceeds as follows.  At each time step
$\Delta t$, the state $|\psi(t)\rangle$ evolves either through a quantum jump with probability $\delta p(t)$ or according to an effective non-Hermitian Hamiltonian dynamics (no-jump evolution). The quantum jump occurs with probability,
\begin{equation}
\delta p(t)
=
\Delta t
\sum_{\mu}
\langle\psi(t)|\hat L_\mu^\dagger \hat L_\mu|\psi(t)\rangle ,
\label{eq:jump_prob}
\end{equation}
When a jump occurs, the channel $\mu$ is selected with probability proportional to
$\delta p_{\mu}(t) = \langle\psi(t)|\hat L_\mu^\dagger \hat L_\mu|\psi(t)\rangle$, and the state is updated accordingly,
\begin{equation}
\ket{\psi(t)}
\rightarrow
\ket{\psi(t+\Delta t)} = \frac{\hat L_\mu\ket{\psi(t)}}{\|\hat L_\mu\ket{\psi(t)}\|}.
\label{eq:jump}
\end{equation}

If no jump occurs, the state evolves with the effective Hamiltonian,
$ \hat H_{\mathrm{eff}} = \hat H - \frac{i}{2} \sum_{\mu} \hat L_\mu^\dagger \hat L_\mu$, as
\begin{equation}
\ket{\psi(t+\Delta t)}
\propto
e^{-i\hat H_{\mathrm{eff}}\Delta t}
\ket{\psi(t)},
\end{equation}
followed by normalization. Repeating this procedure generates a stochastic trajectory describing the open-system dynamics.

In this approach the expectation values of observables are computed along each trajectory and subsequently averaged over an ensemble of independent realizations. Explicitly, for an observable $\hat O$, the trajectory average is given by
\begin{equation}
\langle \hat O(t) \rangle
=
\frac{1}{N_{\mathrm{traj}}}
\sum_{\ell=1}^{N_{\mathrm{traj}}}
\bra{\psi^{[\ell]}(t)} \hat O \ket{\psi^{[\ell]}(t)}.
\label{eq:traj_average}
\end{equation}
where $\ket{\psi^{[\ell]}(t)}$ denotes the $\ell$'th quantum state trajectory. Moreover, as we mentioned, an important advantage of the trajectory formulation is that the global state remains pure along each realization. This allows the entanglement to be evaluated before ensemble averaging, providing a clear distinction between genuine quantum correlations and classical mixing induced by averaging over stochastic trajectories.

\section{Correlation quantifier}
\label{sec:diagnostics}

To characterize the correlations generated by the dissipative dynamics, we employ an information-theoretic diagnostics that probe correlations at different scales. This quantity, evaluated at the level of individual quantum trajectories and subsequently averaged over stochastic realizations, allow us to quantify the inter-ensemble quantum correlations between the two spin ensembles.
Specifically, for a given trajectory $\ell$, the quantum correlations between ensembles $A$ and $B$ are quantified by the von Neumann entropy of the reduced state
\begin{equation}
\hat \rho_A^{[\ell]}(t)
=
\mathrm{Tr}_B
\left[
|\psi^{[\ell]}(t)\rangle\langle\psi^{[\ell]}(t)|
\right],
\label{eq:rhoA}
\end{equation}
from which the trajectory-resolved entanglement entropy follows as
\begin{equation}
S_A^{[\ell]}(t) = - \mathrm{Tr} \left[ \hat \rho_A^{[\ell]}(t)\log_2\hat \rho_A^{[\ell]}(t) \right].
\label{eq:SAtraj}
\end{equation}
The entanglement entropy reported in our analysis is obtained by averaging over trajectories,
\begin{equation}
S_A(t)
=
\frac{1}{N_{\mathrm{traj}}}
\sum_{\ell=1}^{N_{\mathrm{traj}}}
S_A^{[\ell]}(t).
\label{eq:SA}
\end{equation}
In particular, we are often  interested in the long-time dynamics of the entanglement, when it reaches roughly stable plateaus.
 We define these plateaus as the steady state entanglement, given by
\begin{equation}
\label{eq.ss.ent.ltraj}
S_{A,ss}^{[\ell]}
=
\frac{1}{\Delta t}
\int_{t_0}^{t_f}
S_A^{[\ell]}(t)\,dt
\end{equation}
 with $t_{o}$ and $t_f$ chosen within the plateau  time interval. Ideally, one would have $t_f \rightarrow \infty$, but due to numerical constraints a sufficiently large finite integration time is used to guarantee convergence of the computed steady-state entanglement.
Similarly, the average steady state entanglement is defined by,
 \begin{equation}
S_{A,ss} =
\frac{1}{N_{\mathrm{traj}}}
\sum_{\ell=1}^{N_{\mathrm{traj}}}
S_{A,ss}^{[\ell]}.
\end{equation}

We remark that the entanglement entropy has been widely used as a fundamental diagnostic for quantum correlations in many-body systems, with particular utility in non-equilibrium settings where unitary dynamics compete with measurement processes \cite{li_measurement-induced_2025,christopoulosNoisyQuantumDynamics2024,fazioManyBody2024}. A paradigmatic example is the measurement-induced phase transition (MIPT), which occurs in monitored quantum circuits where unitary evolution alternates with projective measurements on a fraction of the system. In this settings, the competition between unitary and measurements drives a sharp transition: usually when the measurement rate is low, the steady state exhibits volume-law entanglement entropy, indicative of a highly entangled phase where quantum information spreads throughout the system; when the measurement rate exceeds a critical threshold, entanglement entropy switches to area-law scaling, reflecting a disentangled phase where measurements repeatedly localize quantum information and suppress long-range correlations. This transition is not captured by traditional equilibrium order parameters but is uniquely diagnosed by the scaling of entanglement entropy across the system, as well as by its fluctuations.

\section{Results}
\label{sec:inter.ensemble}

In this section we explore both the dynamical growth of inter-ensemble correlations, as well as its steady states properties and phase diagram. We also perform a finite size scaling for the correlations.

\subsection{Dynamics}

We begin by analyzing the dynamical generation of correlations between the two ensembles. In all our simulations we consider the ensembles prepared in the same initial uncorrelated product state,
\begin{equation}
\ket{\psi(0)}=\ket{\uparrow}^{\otimes N_A}\otimes\ket{\uparrow}^{\otimes N_B}.
\end{equation}
Therefore, any correlations observed during the evolution must be generated by the dissipative coupling mediated by the collective decay channel.
Unless otherwise stated, all numerical results shown below are obtained for $\omega_A=1.2$, $\omega_B=0.95$, and $\kappa_A=\kappa_B=1.0$, and average over $N_{\mathrm{Traj}} = 1000$ trajectories.

In Fig.~\ref{fig:entanglement_dynamics} we show the time evolution of the inter-ensemble entanglement entropy $S_A(t)$ for several system sizes and representative values of the collective decay strength $\kappa_{AB}$. We observe that as the system evolves, the collective dissipative processes generate correlations between the ensembles, leading to a rapid growth of $S_A(t)$. The growth of correlations is followed by the emergence of a pronounced peak in the entanglement entropy. The height and position of this peak depend on the strength of the collective dissipative coupling $\kappa_{AB}$.
At longer times the correlations partially relax and approach a stationary plateau, corresponding to the steady-state correlations of the system.

Importantly, the qualitative behaviour of the dynamics depends strongly on the strength of the collective dissipative coupling. For weaker coupling ($\kappa_{AB} = 0.1$), correlations grow relatively slowly, whereas for intermediate coupling ($\kappa_{AB} = 0.22$), they are significantly enhanced and reach steady-state entanglement with a higher value. By contrast, stronger collective decay ($\kappa_{AB} = 1$) results in a weakly entangled steady state, with correlations decaying rapidly after the initial transient.  We analyse such observations in more detail below, scrutinising the steady-state phase diagram for correlations and providing a quantitative analysis of the dynamical rates.

\subsection{Steady-state}

\begin{figure}
\centering
\includegraphics[width=0.95\columnwidth]{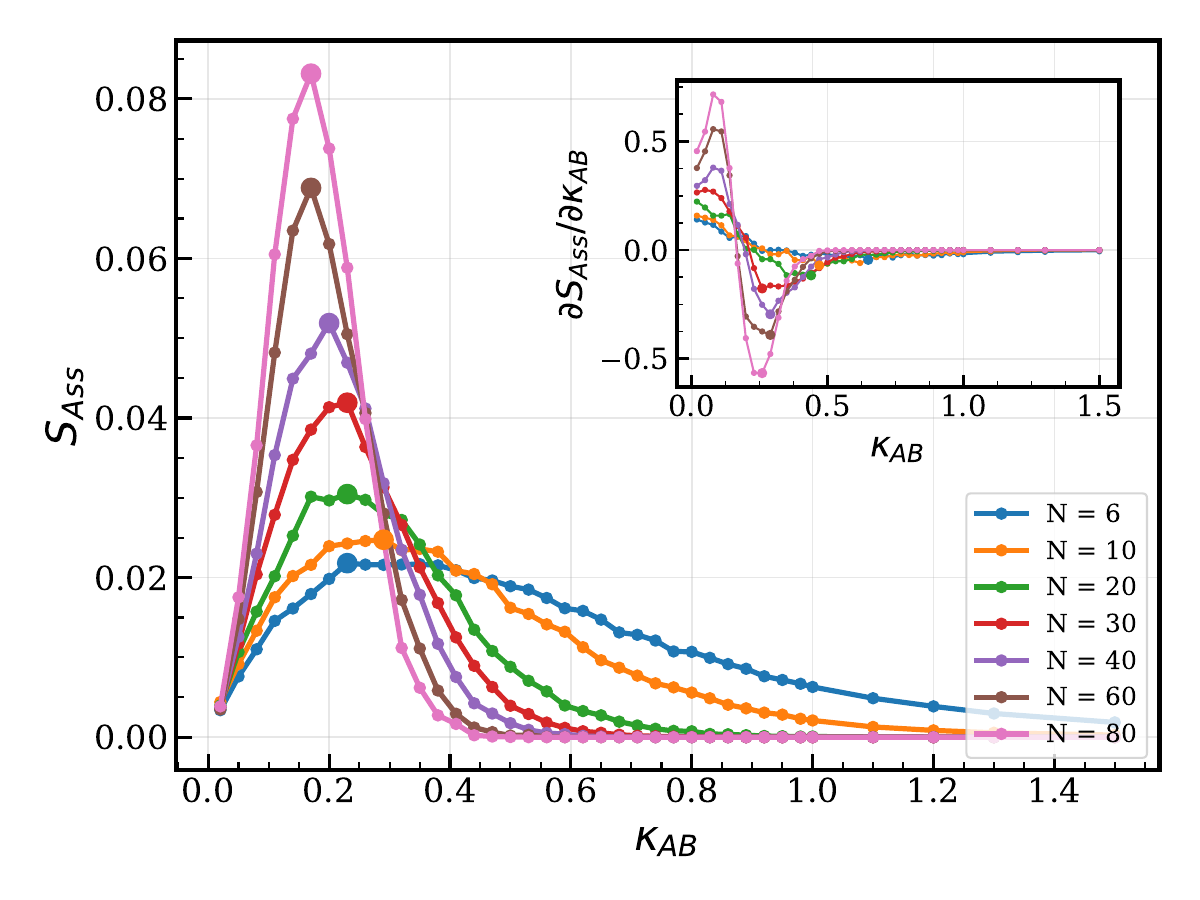}
\caption{
Steady-state average entanglement entropy
$\langle S_A \rangle_{\mathrm{ss}}$ for varying  collective decay
strength $\kappa_{AB}$ and  system sizes. While for small couplings, in the seeded-BTC phase, it tends to increases with system size, for strong couplings, in the static non-seeded phase, it rather decreases with $N$. The steady-state correlations exhibit a pronounced maximum (marked by larger circles) at intermediate coupling.
  In the (inset panel) we show the entanglement susceptibility, $\partial S_A / \partial \kappa_{AB}$, highlighting its steep behavior around the mean-field transition.
  }
\label{fig:phase_diagram}
\end{figure}

In Fig.\eqref{fig:phase_diagram} we show the steady-state inter-ensemble entanglement entropy as a function of the collective decay strength $\kappa_{AB}$ for several system sizes. As the coupling strength increases, the steady-state entanglement grows and reaches a pronounced maximum at intermediate coupling strengths $\kappa_{AB} \sim 0.2$. After that, it rapidly decreases to rather small values.

\begin{figure*}
\centering
\begin{subfigure}{0.32\linewidth}
\centering
\hspace{-10mm}
\includegraphics[width=\linewidth]{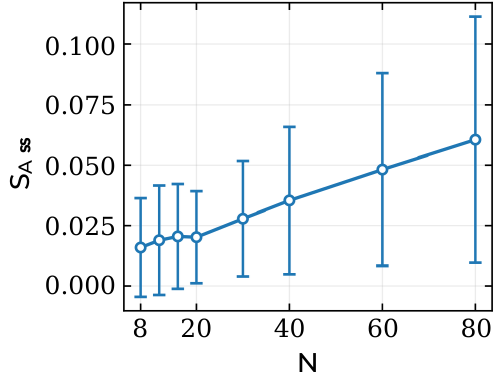}
\caption{\centering $\kappa_{AB}=0.1$}
\end{subfigure}
\hfill
\begin{subfigure}{0.32\linewidth}
\centering
\hspace{-10mm}
\includegraphics[width=\linewidth]{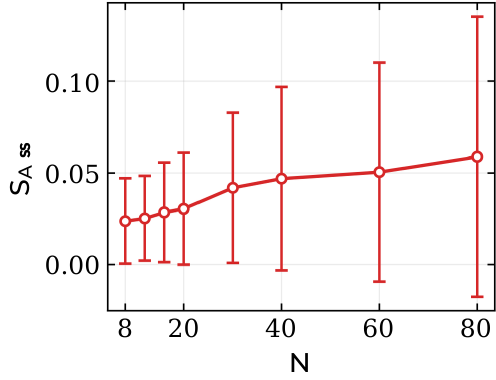}
\caption{\centering $\kappa_{AB}=0.22$}
\end{subfigure}
\hfill
\begin{subfigure}{0.32\linewidth}
\centering
\hspace{-10mm}
\includegraphics[width=\linewidth]{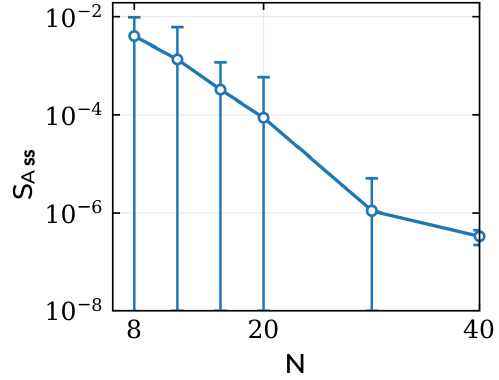}
\caption{\centering $\kappa_{AB}=1.0$}
\end{subfigure}
\caption{
Finite-size scaling analysis of the steady-state inter-ensemble entanglement entropy for representative values of the collective decay strength $\kappa_{AB}$. The regime of weak collective decay exhibits an increasing entanglement with system size, while for strong collective decays the entanglement is exponentially suppressed with the system size.
The color bars represent the trajectory fluctuations of the entanglement entropy, which either grows with the system size in the seeded BTC or decreases in the non-seeded phase.
}
\label{fig:entanglement_scaling}
\end{figure*}

There are several interesting properties worth mentioning. As previously mentioned, a mean-field analysis of the average magnetisation predicts a phase transition from a seeded-BTC phase to a static non-BTC phase at the critical coupling as given by Eq.\eqref{eq:Gamma_crit}.  In our case, the mean-field analysis of the magnetization predicts a critical coupling $\kappa_{\mathrm{AB,crit}} \approx 0.235$. The peak of the steady-state entanglement entropy is located near this value, though not exactly at it; as the system size increases, the peak position shifts to slightly smaller couplings. Notably, the derivative of the steady-state entanglement with respect to the dissipative coupling, $\partial S_A / \partial \kappa_{AB}$, becomes increasingly steep in the vicinity of the mean-field critical value - see Fig.\eqref{fig:phase_diagram}-(inset panel). This suggests that the derivative may provide a more sensitive indicator of the transition than the entanglement itself, and its behavior appears consistent with the critical coupling scale derived from magnetization observables. While a more detailed finite-size scaling analysis would be required to precisely locate the critical point from entanglement data, our results indicate that the entanglement susceptibility captures the same underlying phase boundary.

Moreover, we observe that while the entanglement increases with system size in the seeded BTC phase ($\kappa_{AB} \lesssim \kappa_{\mathrm{AB,crit}}$), the opposite occurs in the static phase, where the entanglement decreases with system size,therefore exhibiting a MIPT.
To further characterize these regimes, we perform a finite-size scaling analysis of the entanglement in the two phases, as shown in
Fig.\eqref{fig:entanglement_scaling}. In the seeded-BTC phase, the  entanglement scales with the system size with $ S_{A,ss}  \propto N$, as shown in Fig.\eqref{fig:entanglement_scaling}-(a). In contrast, it exhibits an exponential decay in the static phase, with $  S_{A,ss}  \propto e^{-g N}$ (see Fig.\eqref{fig:entanglement_scaling}-(c)), where $g>0$ depends on the system parameters. Therefore, despite still exchanging information and correlations with each other for an intermediate time, the two ensembles tend to become completely decoupled over long timescales and for large system sizes. This is in contrast to the highly correlated steady-state ensembles in the seeded-BTC phase. Close to the critical transition, nevertheless, it is difficult to determine its scaling behaviour. For our available system sizes, it exhibits slow growth with system size (Fig.\eqref{fig:entanglement_scaling}-(b)), but we cannot precisely determine its scaling law (mainly due to our constraint of analysing systems of a relatively small sizes).

Notably the trajectory fluctuations of the entanglement entropy
 exhibit a markedly different behavior across the two phases - see color bars in Fig.\eqref{fig:entanglement_scaling}.  This is defined as the standard deviation,
\begin{eqnarray}
\delta S_{A,ss} = \sqrt{
\frac{1}{N_{\mathrm{traj}}-1}
\sum_{\ell=1}^{N_{\mathrm{traj}}}
\left(
S_{A,ss}^{[\ell]}
-
S_{A,ss}.
\right)^2 }
\end{eqnarray}
In the seeded-BTC phase ($\kappa_{AB} \lesssim \kappa_{\mathrm{AB,crit}}$), the fluctuations tend to grow
with system size $N$, which we attribute to the persistent dynamics
supported in this phase: due to the persistent dynamics
individual trajectories explore larger portions of the full Hilbert space, therefore in a broader range of entanglement values leading
to larger variance across realizations. In contrast, in the static non-seeded phase
($\kappa_{AB} \gtrsim \kappa_{\mathrm{AB,crit}}$), the fluctuations decrease with
$N$, consistent with the rapid equilibration of the system toward its steady state, where trajectories become increasingly concentrated around a single entanglement value.

\subsection{Correlation growth and decay rates}

To further characterize the dynamical generation and relaxation of correlations, we analyze effective rates associated with the initial transient growth and decay of the inter-ensemble entanglement entropy. We extract these quantities directly from the time dependence of the entanglement entropy $S_A(t)$ using a global slope estimator, which provides a measures of the overall rate of change in the dynamics.

Specifically, the growth rate $\Gamma_{\mathrm{grow}}$ measures the rate at which correlations are injected into the system during the initial stage of the evolution, defined as the slope between the initial time and the first maximum of the entanglement entropy,
\begin{equation}
\Gamma_{\mathrm{grow}}
=
\frac{S_A(t_{\mathrm{max}}) - S_A(0)}{t_{\mathrm{max}}},
\label{eq:growth_rate}
\end{equation}
where $t_{\mathrm{max}}$ denotes the time at which $S_A(t)$ reaches its first maximum.
Similarly, the decay rate $\Gamma_{\mathrm{decay}}$
  characterizes the relaxation of correlations after this peak, defined as  the slope between the first maximum and the first subsequent minimum,
\begin{equation}
\Gamma_{\mathrm{decay}}
=
\frac{S_A(t_{\mathrm{min}}) - S_A(t_{\mathrm{max}})}{t_{\mathrm{min}} - t_{\mathrm{max}}},
\label{eq:decay_rate}
\end{equation}
where $t_{\mathrm{min}}$ is the time of the first minimum following the initial peak. We note that in the strong-coupling regime
($\kappa_{AB} \gtrsim \kappa_{\mathrm{AB,crit}}$), the entanglement entropy does not exhibit a pronounced secondary minimum after the initial peak. Instead, it relaxes monotonically toward the steady-state value. In this regime, we therefore define $t_{\mathrm{min}}$ as the time at which $S_A(t)$ first reaches within
$1\%$ of its long-time steady-state value $\langle S_A \rangle_{\mathrm{ss}}$, providing a consistent operational definition of the decay rate across all coupling
regimes.

\begin{figure}
\includegraphics[width=0.9\linewidth]{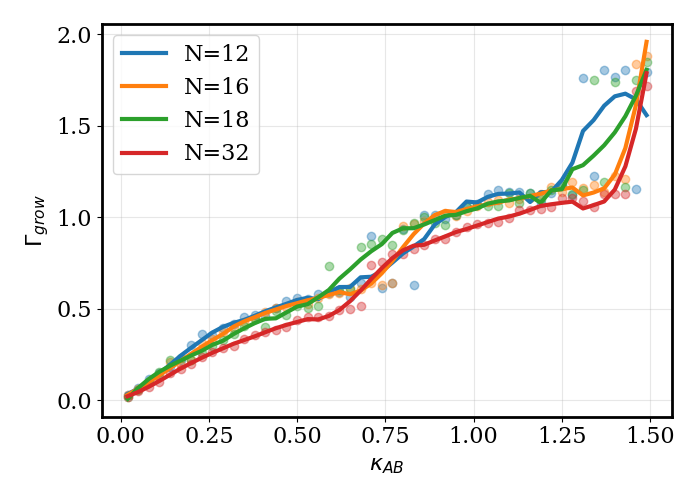}
\includegraphics[width=0.9\linewidth]{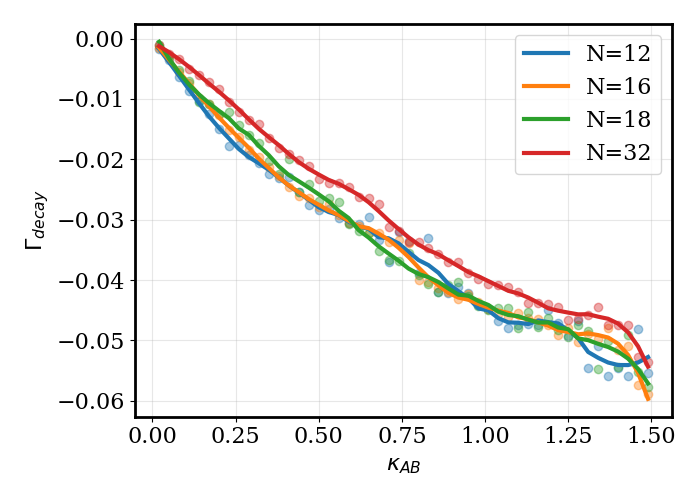}
\caption{
Effective dynamical rates extracted from the entanglement dynamics $S_A(t)$. Solid lines show a Savitzky–Golay filter smoothing of the data to highlight overall trends.
(a) Growth rate of correlations ($\Gamma_{\mathrm{grow}}$ - Eq.\eqref{eq:growth_rate}).  (b) Decay rate  ($\Gamma_{\mathrm{decay}}$ - Eq.\eqref{eq:decay_rate}).
Both  rates increases in absolute value with the collective decay strength $\kappa_{AB}$, indicating a faster exchange of information and correlations directly dependent on the the collective dissipation.
}
\label{fig:injection_decay}
\end{figure}

Figure~\ref{fig:injection_decay} shows the dependence of these quantities on the collective decay strength $\kappa_{AB}$ for different system sizes $N$.
We observe that the growth rate increases with $\kappa_{AB}$, indicating that stronger collective dissipation enhances the speed at which correlations are generated between the ensembles.
 Similarly, the decay rate becomes increasingly negative, reflecting the enhanced  dissipative suppression of correlations at larger coupling strengths.  The overall behavior thus indicates that correlations are rapidly suppressed at early times after the peak, followed by a slower relaxation at later times.
At all coupling strengths, both rates show a relatively
weak dependence on system size $N$, suggesting that the dynamical timescales governing correlation growth and decay are primarily set by the dissipative parameters $\kappa_{AB}$, $\kappa_\alpha$, and $\omega_\alpha$, rather than by the system size.

\section{Conclusions}
\label{sec.conclusion}

In this work, we have investigated, using a quantum-trajectory framework, how dissipative seeding generates inter-ensemble correlations in a system of two spin ensembles. One ensemble exhibits boundary time-crystal (BTC) behavior, while the other is initially static. Coupling them via a shared collective decay channel allows the time-crystalline order to seed into the static partner, profoundly affecting the quantum correlations between the two. Our results reveal a MIPT in the steady-state entanglement and its fluctuations, driven by the seeding mechanism. Overall, our findings establish dissipative seeding as a mechanism not only for synchronizing macroscopic observables (such as bare magnetization) but also for controlling the scaling of genuine quantum correlations in open many-body systems.

Beyond these central findings, our work opens several avenues for further investigation. First, the trajectory-resolved approach adopted here naturally extends to systems with more than two ensembles. One could envision a network of spin collectives where a single BTC node seeds multiple static neighbors, potentially giving rise to richer entanglement architectures.  Second, the observed entanglement trajectories and MIPT occur without the need for postselection, a feature that distinguishes this model from conventional monitored circuits and makes it particularly attractive in ongoing experiments with cold atoms and trapped ions. This raises the interesting possibility of using time-crystalline order as a tunable resource for generating genuine many-body entanglement in steady-state open systems. A more detailed analysis of such a control, with the dependence  on  further experimental imperfections and interactions in the model implementation seems interesting directions.  Third, our analysis focused on symmetric ensembles of equal size. An interesting perspective is to explore asymmetric configurations, where the seeding directionality could possibly be controlled or even reversed by tuning the relative sizes or local dissipation rates.

\begin{acknowledgments}
 We acknowledge financial support from the Brazilian funding agencies CAPES, CNPq (308637/2022-4), FAPERJ (No. E-26/210.236/2024, and No.E-26/204.340/2025), and by the Serrapilheira Institute (grant number Serra 2211-42166).
\end{acknowledgments}

\bibliographystyle{apsrev4-2}
\bibliography{references}

@article{PhysRevA.109.L050203,
  title = {Floquet time crystals as quantum sensors of ac fields},
  author = {Iemini, Fernando and Fazio, Rosario and Sanpera, Anna},
  journal = {Phys. Rev. A},
  volume = {109},
  issue = {5},
  pages = {L050203},
  numpages = {6},
  year = {2024},
  month = {May},
  publisher = {American Physical Society},
  doi = {10.1103/PhysRevA.109.L050203},
  url = {https://link.aps.org/doi/10.1103/PhysRevA.109.L050203}
}

@article{PhysRevB.111.024315,
  title = {Prethermal Floquet time crystals in chiral multiferroic chains and applications as quantum sensors of AC fields},
  author = {Shukla, Rohit Kumar and Chotorlishvili, Levan and Mishra, Sunil K. and Iemini, Fernando},
  journal = {Phys. Rev. B},
  volume = {111},
  issue = {2},
  pages = {024315},
  numpages = {10},
  year = {2025},
  month = {Jan},
  publisher = {American Physical Society},
  doi = {10.1103/PhysRevB.111.024315},
  url = {https://link.aps.org/doi/10.1103/PhysRevB.111.024315}
}

@article{7m63-lnb8,
  title = {Discrete time crystal for periodic-field sensing with quantum-enhanced precision},
  author = {Yousefjani, Rozhin and Al-Kuwari, Saif and Bayat, Abolfazl},
  journal = {Phys. Rev. Appl.},
  volume = {24},
  issue = {5},
  pages = {054047},
  numpages = {14},
  year = {2025},
  month = {Nov},
  publisher = {American Physical Society},
  doi = {10.1103/7m63-lnb8},
  url = {https://link.aps.org/doi/10.1103/7m63-lnb8}
}

@misc{manya2026optimalobservablesquantumenhancedsensing,
      title={Optimal observables for quantum-enhanced sensing and applications in a Floquet time crystal sensor}, 
      author={M. A. Manya and Andrei Tsypilnikov and Fernando Iemini},
      year={2026},
      eprint={2606.26248},
      archivePrefix={arXiv},
      primaryClass={quant-ph},
      url={https://arxiv.org/abs/2606.26248}, 
}

@article{9h67-kqyz,
  title = {Exact analysis of ac sensors based on Floquet time crystals},
  author = {Tsypilnikov, Andrei and Fibger, Matheus and Iemini, Fernando},
  journal = {Phys. Rev. A},
  volume = {113},
  issue = {2},
  pages = {022620},
  numpages = {18},
  year = {2026},
  month = {Feb},
  publisher = {American Physical Society},
  doi = {10.1103/9h67-kqyz},
  url = {https://link.aps.org/doi/10.1103/9h67-kqyz}
}

@misc{cabot2026parameterestimationonetwotime,
      title={Parameter estimation with one- and two-time measurements on the emission field of the boundary time crystal}, 
      author={Albert Cabot and Federico Carollo and Igor Lesanovsky},
      year={2026},
      eprint={2503.21753},
      archivePrefix={arXiv},
      primaryClass={quant-ph},
      url={https://arxiv.org/abs/2503.21753}, 
}

@article{montenegroQuantum2023,
	title = {Quantum metrology with boundary time crystals},
	volume = {6},
	copyright = {2023 The Author(s)},
	issn = {2399-3650},
	url = {https://www.nature.com/articles/s42005-023-01423-6},
	doi = {10.1038/s42005-023-01423-6},
	language = {en},
	number = {1},
	urldate = {2024-08-16},
	journal = {Communications Physics},
	publisher = {Nature Publishing Group},
	author = {Montenegro, Victor and Genoni, Marco G. and Bayat, Abolfazl and Paris, Matteo G. A.},
	month = oct,
	year = {2023},
	pages = {1--9}
}

@article{MONTENEGRO20251,
title = {Review: Quantum metrology and sensing with many-body systems},
journal = {Physics Reports},
volume = {1134},
pages = {1-62},
year = {2025},
note = {Review: Quantum metrology and sensing with many-body systems},
issn = {0370-1573},
doi = {https://doi.org/10.1016/j.physrep.2025.05.005},
url = {https://www.sciencedirect.com/science/article/pii/S0370157325001565},
author = {Victor Montenegro and Chiranjib Mukhopadhyay and Rozhin Yousefjani and Saubhik Sarkar and Utkarsh Mishra and Matteo G.A. Paris and Abolfazl Bayat}
}

@misc{liEmergent2025,
	title = {Emergent deterministic entanglement dynamics in monitored infinite-range bosonic systems},
	url = {http://arxiv.org/abs/2506.18624},
	doi = {10.48550/arXiv.2506.18624},
	urldate = {2025-06-27},
	publisher = {arXiv},
	author = {Li, Zejian and Delmonte, Anna and Fazio, Rosario},
	month = jun,
	year = {2025},
	note = {arXiv:2506.18624 [quant-ph]}
}

@article{delmonteMeasurementinduced2025,
	title = {Measurement-induced phase transitions in monitored infinite-range interacting systems},
	volume = {7},
	issn = {2643-1564},
	url = {https://link.aps.org/doi/10.1103/PhysRevResearch.7.023082},
	doi = {10.1103/PhysRevResearch.7.023082},
	language = {en},
	number = {2},
	urldate = {2025-07-03},
	journal = {Physical Review Research},
	author = {Delmonte, Anna and Li, Zejian and Passarelli, Gianluca and Song, Eric Yilun and Barberena, Diego and Rey, Ana Maria and Fazio, Rosario},
	year = {2025},
	pages = {023082}
}

@article{SciPostPhys.18.3.100,
	title = {Boundary time crystals as AC sensors: Enhancements and constraints},
	pages = {100},
	author = {Gribben, Dominic and Sanpera, Anna and Fazio, Rosario and Marino, Jamir and Iemini, Fernando},
	journal = {SciPost Phys.},
	volume = {18},
	year = {2025},
	publisher = {SciPost},
	doi = {10.21468/SciPostPhys.18.3.100},
	url = {https://scipost.org/10.21468/SciPostPhys.18.3.100}
}

@article{PhysRevLett.132.050801,
  title = {Continuous Sensing and Parameter Estimation with the Boundary Time Crystal},
  author = {Cabot, Albert and Carollo, Federico and Lesanovsky, Igor},
  journal = {Phys. Rev. Lett.},
  volume = {132},
  issue = {5},
  pages = {050801},
  numpages = {6},
  year = {2024},
  month = {Jan},
  publisher = {American Physical Society},
  doi = {10.1103/PhysRevLett.132.050801},
  url = {https://link.aps.org/doi/10.1103/PhysRevLett.132.050801}
}

@misc{lee2025timescalessqueezingheisenbergscalings,
      title={Timescales, Squeezing and Heisenberg Scalings in Many-Body Continuous Sensing}, 
      author={Gideon Lee and Ron Belyansky and Liang Jiang and Aashish A. Clerk},
      year={2025},
      eprint={2505.04591},
      archivePrefix={arXiv},
      primaryClass={quant-ph},
      url={https://arxiv.org/abs/2505.04591}, 
}

@article{Li_2025,
   title={Exact steady state of the quantum van der Pol oscillator: Critical phenomena and enhanced metrology},
   volume={112},
   ISSN={2469-9934},
   url={http://dx.doi.org/10.1103/zybb-vxfz},
   DOI={10.1103/zybb-vxfz},
   number={2},
   journal={Physical Review A},
   publisher={American Physical Society (APS)},
   author={Li, Yaohua and Zhang, Xuanchen and Liu, Yong-Chun},
   year={2025},
   month=Aug }

@article{5gh9nmv8,
  title = {Designing Open Quantum Systems for Enabling Quantum-Enhanced Sensing through Classical Measurements},
  author = {Mattes, Robert and Cabot, Albert and Carollo, Federico and Lesanovsky, Igor},
  journal = {Phys. Rev. Lett.},
  volume = {135},
  issue = {23},
  pages = {230402},
  numpages = {8},
  year = {2025},
  month = {Dec},
  publisher = {American Physical Society},
  doi = {10.1103/5gh9-nmv8},
  url = {https://link.aps.org/doi/10.1103/5gh9-nmv8}
}

@misc{midha2025metrologyopenquantumsystems,
      title={Metrology of open quantum systems from emitted radiation}, 
      author={Siddhant Midha and Sarang Gopalakrishnan},
      year={2025},
      eprint={2504.13815},
      archivePrefix={arXiv},
      primaryClass={quant-ph},
      url={https://arxiv.org/abs/2504.13815}, 
}

@misc{oconnor2026attainingquantumsensingenhancement,
      title={Attaining Quantum Sensing Enhancement from Monitored Dissipative Time Crystals}, 
      author={Eoin O'Connor and Victor Montenegro and Francesco Albarelli and Matteo G. A. Paris and Abolfazl Bayat and Marco G. Genoni},
      year={2026},
      eprint={2508.15448},
      archivePrefix={arXiv},
      primaryClass={quant-ph},
      url={https://arxiv.org/abs/2508.15448}, 
}

@article{Pavlov_2023,
   title={Quantum metrology with critical driven-dissipative collective spin system},
   volume={98},
   ISSN={1402-4896},
   url={http://dx.doi.org/10.1088/1402-4896/ace99f},
   DOI={10.1088/1402-4896/ace99f},
   number={9},
   journal={Physica Scripta},
   publisher={IOP Publishing},
   author={Pavlov, Venelin P and Porras, Diego and Ivanov, Peter A},
   year={2023},
   month=Aug, pages={095103} }

@article{arumugamElectricfield2025,
	title = {Electric-field sensing with driven-dissipative time crystals in room-temperature {Rydberg} vapor},
	volume = {15},
	copyright = {2025 The Author(s)},
	issn = {2045-2322},
	url = {https://www.nature.com/articles/s41598-025-97560-9},
	doi = {10.1038/s41598-025-97560-9},
	language = {en},
	number = {1},
	urldate = {2025-06-12},
	journal = {Scientific Reports},
	publisher = {Nature Publishing Group},
	author = {Arumugam, Darmindra},
	month = apr,
	year = {2025},
	pages = {13446}
}

@misc{singh2025quantumthermodynamicslimitcycle,
      title={Quantum Thermodynamics on a limit cycle}, 
      author={Varinder Singh and Euijoon Kwon and G J Milburn},
      year={2025},
      eprint={2503.12118},
      archivePrefix={arXiv},
      primaryClass={quant-ph},
      url={https://arxiv.org/abs/2503.12118}, 
}

@article{dj21gmdj,
  title = {Quantum Time Crystal Clock and Its Performance},
  author = {Viotti, Ludmila and Huber, Marcus and Fazio, Rosario and Manzano, Gonzalo},
  journal = {Phys. Rev. Lett.},
  volume = {136},
  issue = {11},
  pages = {110401},
  numpages = {8},
  year = {2026},
  month = {Mar},
  publisher = {American Physical Society},
  doi = {10.1103/dj21-gmdj},
  url = {https://link.aps.org/doi/10.1103/dj21-gmdj}
}

@article{Carollo_2024,
doi = {10.1088/2058-9565/ad3f42},
url = {https://doi.org/10.1088/2058-9565/ad3f42},
year = {2024},
month = {may},
publisher = {IOP Publishing},
volume = {9},
number = {3},
pages = {035024},
author = {Carollo, Federico and Lesanovsky, Igor and Antezza, Mauro and De Chiara, Gabriele},
title = {Quantum thermodynamics of boundary time-crystals},
journal = {Quantum Science and Technology},
}

@article{Paulino_2026,
doi = {10.1088/2058-9565/ae186c},
url = {https://doi.org/10.1088/2058-9565/ae186c},
year = {2025},
month = {nov},
publisher = {IOP Publishing},
volume = {11},
number = {1},
pages = {015003},
author = {Paulino, Paulo J and Cabot, Albert and De Chiara, Gabriele and Antezza, Mauro and Lesanovsky, Igor and Carollo, Federico},
title = {Thermodynamics of coupled time crystals with an application to energy storage},
journal = {Quantum Science and Technology},
}

@article{zheludevTime2024,
	title = {Time crystals for photonics and timetronics},
	volume = {18},
	issn = {1749-4885, 1749-4893},
	url = {https://www.nature.com/articles/s41566-024-01557-1},
	doi = {10.1038/s41566-024-01557-1},
	language = {en},
	number = {11},
	urldate = {2025-06-11},
	journal = {Nature Photonics},
	author = {Zheludev, Nikolay I.},
	month = nov,
	year = {2024},
	pages = {1123--1125}
}

@article{li_measurement-induced_2025,
	title = {Measurement-induced entanglement phase transition in free fermion systems},
	volume = {37},
	issn = {0953-8984},
	url = {https://doi.org/10.1088/1361-648X/ade7e5},
	doi = {10.1088/1361-648X/ade7e5},
	language = {en},
	number = {27},
	urldate = {2026-06-15},
	journal = {Journal of Physics: Condensed Matter},
	publisher = {IOP Publishing},
	author = {Li, Han-Ze and Zhong, Jian-Xin and Yu, Xue-Jia},
	month = jul,
	year = {2025},
	pages = {273002},
}

@misc{christopoulosNoisyQuantumDynamics2024,
      title={Cahier de l'Institut Pascal: Noisy Quantum Dynamics and Measurement-Induced Phase Transitions}, 
      author={Alexios Christopoulos and Alessandro Santini and Guido Giachetti},
      year={2024},
      eprint={2409.06310},
      archivePrefix={arXiv},
      primaryClass={cond-mat.stat-mech},
      url={https://arxiv.org/abs/2409.06310}, 
}

@article{fazioManyBody2024,
	title = {Many-body open quantum systems},
	pages = {99},
	author = {Fazio, Rosario and Keeling, Jonathan and Mazza, Leonardo and Schirò, Marco},
	journal = {SciPost Phys. Lect. Notes},
	year = {2025},
	publisher = {SciPost},
	doi = {10.21468/SciPostPhysLectNotes.99},
	url = {https://scipost.org/10.21468/SciPostPhysLectNotes.99}
}

@article{Diehl2008,
  author = {Diehl, S. and Micheli, A. and Kantian, A. and Kraus, B. and B{\"u}chler, H. P. and Zoller, P.},
  title = {Quantum states and phases in driven open quantum systems with cold atoms},
  journal = {Nature Physics},
  year = {2008},
  volume = {4},
  number = {11},
  pages = {878--883},
  doi = {10.1038/nphys1073}
}

@article{Verstraete2009,
  author = {Verstraete, Frank and Wolf, Michael M. and Cirac, J. Ignacio},
  title = {Quantum computation and quantum-state engineering driven by dissipation},
  journal = {Nature Physics},
  year = {2009},
  volume = {5},
  number = {9},
  pages = {633--636},
  doi = {10.1038/nphys1342}
}

@article{Daley2014,
  author = {Daley, Andrew J.},
  title = {Quantum trajectories and open many-body quantum systems},
  journal = {Advances in Physics},
  year = {2014},
  volume = {63},
  number = {2},
  pages = {77--149},
  doi = {10.1080/00018732.2014.933502}
}

@article{Sieberer2016,
  author = {Sieberer, L. M. and Buchhold, M. and Diehl, S.},
  title = {Keldysh field theory for driven open quantum systems},
  journal = {Reports on Progress in Physics},
  year = {2016},
  volume = {79},
  number = {9},
  pages = {096001},
  doi = {10.1088/0034-4885/79/9/096001}
}

@article{Dalibard1992,
  author = {Dalibard, Jean and Castin, Yvan and M{\o}lmer, Klaus},
  title = {Wave-function approach to dissipative processes in quantum optics},
  journal = {Physical Review Letters},
  year = {1992},
  volume = {68},
  number = {5},
  pages = {580--583},
  doi = {10.1103/PhysRevLett.68.580}
}

@article{Plenio1998,
  author = {Plenio, M. B. and Knight, P. L.},
  title = {The quantum-jump approach to dissipative dynamics in quantum optics},
  journal = {Reviews of Modern Physics},
  year = {1998},
  volume = {70},
  number = {1},
  pages = {101--144},
  doi = {10.1103/RevModPhys.70.101}
}

@article{Hajdusek2022,
  author = {Hajdu{\v{s}}ek, Michal and Solanki, Parvinder and Fazio, Rosario and Vinjanampathy, Sai},
  title = {Seeding Crystallization in Time},
  journal = {Physical Review Letters},
  year = {2022},
  volume = {128},
  number = {8},
  pages = {080603},
  doi = {10.1103/PhysRevLett.128.080603}
}

@article{Iemini2018,
  author = {Iemini, F. and Russomanno, A. and Keeling, J. and Schir{\`o}, M. and Dalmonte, M. and Fazio, R.},
  title = {Boundary Time Crystals},
  journal = {Physical Review Letters},
  year = {2018},
  volume = {121},
  number = {3},
  pages = {035301},
  doi = {10.1103/PhysRevLett.121.035301}
}

@article{Krishna2023,
  author = {Krishna, Midhun and Solanki, Parvinder and Hajdu{\v{s}}ek, Michal and Vinjanampathy, Sai},
  title = {Measurement-Induced Continuous Time Crystals},
  journal = {Physical Review Letters},
  year = {2023},
  volume = {130},
  number = {15},
  pages = {150401},
  doi = {10.1103/PhysRevLett.130.150401}
}

@article{Kessler2021,
  author = {Ke{\ss}ler, Hans and Kongkhambut, Phatthamon and Georges, Christoph and Mathey, Ludwig and Cosme, Jayson G. and Hemmerich, Andreas},
  title = {Observation of a Dissipative Time Crystal},
  journal = {Physical Review Letters},
  year = {2021},
  volume = {127},
  number = {4},
  pages = {043602},
  doi = {10.1103/PhysRevLett.127.043602}
}

@article{Chen2023,
  author = {Chen, Yu-Hui and Zhang, Xiangdong},
  title = {Realization of an inherent time crystal in a dissipative many-body system},
  journal = {Nature Communications},
  year = {2023},
  volume = {14},
  pages = {6161},
  doi = {10.1038/s41467-023-41905-3}
}

@article{Wu2024,
  author = {Wu, Xiaoling and Wang, Zhuqing and Yang, Fan and Gao, Ruochen and Liang, Chao and Tey, Meng Khoon and Li, Xiangliang and Pohl, Thomas and You, Li},
  title = {Dissipative time crystal in a strongly interacting Rydberg gas},
  journal = {Nature Physics},
  year = {2024},
  volume = {20},
  number = {9},
  pages = {1389--1394},
  doi = {10.1038/s41567-024-02542-9}
}

@article{Yang2025,
  author = {Yang, Shu and Wang, Zeqing and Fu, Libin and Jie, Jianwen},
  title = {Emergent continuous time crystal in dissipative quantum spin system without driving},
  journal = {Communications Physics},
  year = {2025},
  volume = {8},
  pages = {114},
  doi = {10.1038/s42005-025-02040-1}
}

@article{Huang2025,
  author = {Huang, Ying and Wang, Tishuo and Yin, Haochuan and Jiang, Min and Luo, Zhihuang and Peng, Xinhua},
  title = {Observation of continuous time crystals and quasi-crystals in spin gases},
  journal = {Nature Communications},
  year = {2025},
  volume = {16},
  pages = {9375},
  doi = {10.1038/s41467-025-64413-y}
}

@article{Wang2025,
  author = {Wang, Zhuqing and Gao, Ruochen and Wu, Xiaoling and Bu{\v{c}}a, Berislav and M{\o}lmer, Klaus and You, Li and Yang, Fan},
  title = {Boundary Time Crystals Induced by Local Dissipation and Long-Range Interactions},
  journal = {Physical Review Letters},
  year = {2025},
  volume = {135},
  number = {23},
  pages = {230401},
  doi = {10.1103/jhd4-1khw}
}

@article{Arumugam2026,
  author = {Arumugam, Darmindra},
  title = {Injection locking of Rydberg dissipative time crystals},
  journal = {Communications Physics},
  year = {2026},
  doi = {10.1038/s42005-026-02585-9},
  note = {Published online 18 March 2026}
}

@article{Lourenco2022,
  author = {Louren{\c{c}}o, Ant{\^o}nio C. and dos Prazeres, Luis Fernando and Maciel, Thiago O. and Iemini, Fernando and Duzzioni, Eduardo I.},
  title = {Genuine multipartite correlations in a boundary time crystal},
  journal = {Physical Review B},
  year = {2022},
  volume = {105},
  number = {13},
  pages = {134422},
  doi = {10.1103/PhysRevB.105.134422}
}

@article{Souza2023,
  author = {da Silva Souza, Leonardo and dos Prazeres, Luis Fernando and Iemini, Fernando},
  title = {Sufficient Condition for Gapless Spin-Boson Lindbladians, and Its Connection to Dissipative Time Crystals},
  journal = {Physical Review Letters},
  year = {2023},
  volume = {130},
  number = {18},
  pages = {180401},
  doi = {10.1103/PhysRevLett.130.180401}
}

@article{Carollo2022,
  author = {Carollo, Federico and Lesanovsky, Igor},
  title = {Exact solution of a boundary time-crystal phase transition: Time-translation symmetry breaking and non-Markovian dynamics of correlations},
  journal = {Physical Review A},
  year = {2022},
  volume = {105},
  number = {4},
  pages = {L040202},
  doi = {10.1103/PhysRevA.105.L040202}
}

@article{Passarelli2024,
  author = {Passarelli, Gianluca and Turkeshi, Xhek and Russomanno, Angelo and Lucignano, Procolo and Schir{\`o}, Marco and Fazio, Rosario},
  title = {Many-Body Dynamics in Monitored Atomic Gases without Postselection Barrier},
  journal = {Physical Review Letters},
  year = {2024},
  volume = {132},
  number = {16},
  pages = {163401},
  doi = {10.1103/PhysRevLett.132.163401}
}

@article{LeGal2024,
  author = {Le Gal, Youenn and Turkeshi, Xhek and Schir{\`o}, Marco},
  title = {Entanglement Dynamics in Monitored Systems and the Role of Quantum Jumps},
  journal = {PRX Quantum},
  year = {2024},
  volume = {5},
  number = {3},
  pages = {030329},
  doi = {10.1103/PRXQuantum.5.030329}
}

@article{Shammah2018,
  author = {Shammah, Nathan and Ahmed, Shahnawaz and Lambert, Neill and De Liberato, Simone and Nori, Franco},
  title = {Open quantum systems with local and collective incoherent processes: Efficient numerical simulations using permutational invariance},
  journal = {Physical Review A},
  year = {2018},
  volume = {98},
  number = {6},
  pages = {063815},
  doi = {10.1103/PhysRevA.98.063815}
}

@article{Iemini2024,
  author = {Iemini, Fernando and Chang, Darrick and Marino, Jamir},
  title = {Dynamics of inhomogeneous spin ensembles with all-to-all interactions: Breaking permutational invariance},
  journal = {Physical Review A},
  year = {2024},
  volume = {109},
  number = {3},
  pages = {032204},
  doi = {10.1103/PhysRevA.109.032204}
}
\end{document}